\documentclass[usenatbib]{mn2e}
\usepackage{epsfig}
\begin{document}

\title[Peculiar Velocities into the Next Generation]{Peculiar Velocities into the Next Generation: Cosmological Parameters From Large Surveys without Bias from Nonlinear Structure}

\author[A Abate et al.]{Alexandra Abate$^1$, Sarah Bridle$^1$,  Luis F. A. Teodoro$^2$, Michael S. Warren$^3$, 
\newauthor Martin Hendry$^2$\\
\\
$^1$Department of Physics \& Astronomy, University College London, Gower Street, London, WC1E 6BT, UK\\
$^2$Department of Physics \& Astronomy, Kelvin Building, University of Glasgow, G12 8QQ, Scotland UK\\
$^3$Theoretical Division, Los Alamos National Laboratory, Los Alamos, NM 87545, USA}

\maketitle

\begin{abstract}
We investigate methods to best estimate the normalisation of the mass density fluctuation power spectrum ($\sigma_8$) using peculiar velocity data from a survey like the Six degree Field Galaxy Velocity Survey (6dFGSv). We focus on two potential problems (i) biases from nonlinear growth of structure and (ii) the large number of velocities in the survey. Simulations of $\Lambda$CDM-like models are used to test the methods.  We calculate the likelihood from a full covariance matrix of velocities averaged in grid cells. This simultaneously reduces the number of data points and smooths out nonlinearities which tend to dominate on small scales. We show how the averaging can be taken into account in the predictions in a practical way, and show the effect of the choice of cell size. We find that a cell size can be chosen that significantly reduces the nonlinearities without significantly increasing the error bars on cosmological parameters. We compare our results with those from a principal components analysis following \cite{om} and \cite{feldmanom} to select a set of optimal moments constructed from linear combinations of the peculiar velocities that are least sensitive to the nonlinear scales. We conclude that averaging in grid cells performs equally well.  We find that for a survey such as 6dFGSv we can estimate $\sigma_8$ with less than 3$\%$ bias from nonlinearities. The expected error on $\sigma_8$ after marginalising over $\Omega_m$ is approximately $16$ percent.
\end{abstract}

\section{Introduction}
\renewcommand{\thefootnote}{\fnsymbol{footnote}}
\setcounter{footnote}{1}
\footnotetext{E-mail: aabate@star.ucl.ac.uk}

The formation of structure is assumed to have arisen from seed perturbations created by quantum fluctuations of a scalar field during the inflationary era in the first moment of the universe.  These tiny initial fluctuations in the density field were then amplified by gravitational instability into the structure we perceive around us today. This theory is supported by recent detections of the baryon acoustic oscillations in galaxy power spectra \citep{sdssbao,2dfbao}.  Because these initial fluctuations are assumed to be a Gaussian random field, as suggested by inflationary theory \citep{peeblesbook,bbks}, their distribution will be fully characterised by their power spectrum.  On sufficiently large scales these fluctuations remain linear, even up to the present day, therefore we can simply relate the fluctuation power spectrum today to that at early times.

The distribution of galaxies in the universe is not likely to be the same as the distribution of matter, since most of the mass is in the form of indirectly detectable dark matter.  It is known that galaxies of different types cluster differently \citep{dresslerbias,bias}, so they are clearly not completely unbiased tracers of the underlying mass, and this issue is referred to generically as galaxy biasing. 

Due to its very nature, it is difficult to obtain information on the dark matter distribution. In our local universe the most direct method is to observe the velocities of galaxies relative to the Hubble flow, which arise from the gravitational pull of the dark matter. The peculiar velocity field is a useful tool for probing the matter distribution as galaxy \textit{velocities} are likely to be unbiased traces of the matter velocity field, which in turn is simply related to the density field in linear theory.  Since peculiar velocities are a non-local function of the dark matter distribution then analysing the peculiar velocity field provides information on scales larger than the sampled region \citep{hoff} as the velocity at a point is determined by the integral over the matter distribution in a large volume.

In practice peculiar velocities are complicated by several factors. The major one is that on small scales the density field is highly nonlinear; these effects leak into the velocity field and cannot be described analytically.  A method of adequately separating the contribution from small scales compared to that of large scales must be sought, and this is the problem we focus on in this paper. 

Another major factor is the accuracy of the peculiar velocity measurements.  This relies on knowing the distance to the galaxy through the distance-redshift relation which at low redshift is
$cz=H_0d+v_{pec}$, where the redshift $z$ is the measured spectroscopic redshift, $d$ is the distance to the galaxy and $v_{pec}$ is its radial peculiar velocity.  Therefore to measure the peculiar velocity the distance to the galaxy must first be measured, and this is itself a complicated task.  Relying on the correct calibration of distance indicator relations, the calculated distance is a relative distance measure which is strongly subject to a number of biases and also has very large uncertainties of around 20 percent, all of which translates to the peculiar velocity \cite[see][for a review]{S&W}. The simulations used in this paper contain the large statistical error but the effect of additional biases is beyond the scope of this paper.

Velocities are most sensitive to the cosmological parameter  $\sigma_8$, roughly the amplitude of the power spectrum at a scale of 8$h^{-1}$Mpc, which is a measure of how clustered matter in the universe is today\footnote{Strictly, it is the rms fluctuation of density in spheres of radius 8$h^{-1}$Mpc at the present day, in linear theory.}. It is still not well constrained by any cosmological probe.   The Wilkinson Microwave Anisotropy Probe (WMAP) Cosmic Microwave Background (CMB) measurements \citep{wmap} rely on evolving the anisotropies forward to the present to infer the value of $\sigma_8$, because it is defined at the present epoch.  This will depend on many parameters that affect the growth of structure, such as the mass of the neutrino and the amount or type of dark energy. Allowing the mass of the neutrino to vary significantly alters the WMAP constraints on $\sigma_8$ (see Table \ref{table:sig8}). 

Peculiar velocities provide the only way to measure $\sigma_8$ essentially at redshift zero.  Weak lensing, Lyman-$\alpha$ forest and cluster measurements are obtained only at higher redshifts, where dark energy was just starting to dominate. See Table~\ref{table:sig8} for some current constraints on $\sigma_8$. The range of values could possibly lie anywhere between 0.5 and 1.  The differences in  the values of $\sigma_8$ will of course be in part due to the fact that they are from different experiments, and different parameters have been marginalised over. Pinning down an accurate value of $\sigma_8$ today could help discriminate between different models that affect clustering and the growth of large scale structure such as dark energy models and modified gravity.

\begin{table}
\caption{Some recent 68\% confidence limits for $\sigma_8$ using various cosmological probes;: cosmic microwave background (CMB), weak lensing (WL), Lyman-$\alpha$ forest (Ly$\alpha$), cluster measurements (CL) and supernovae (SN).  This shows that $\sigma_8$ could reasonably lie anywhere in the range 0.5 to 1.}
\begin{center}
\begin{tabular}[b]{|l|l|l|r|} 
\hline 
Authors & Probe & $\sigma_8$ result & Ref\\ 
\hline 
WMAP3 ($\Lambda$CDM) & CMB & 0.76$\pm$0.05 & 1\\
WMAP3 ($\Lambda$CDM+$m_\nu$) & CMB & 0.56$\pm$0.10 & 2\\
Rozo et al. 2007& CL & 0.92$\pm$0.10& 3\\
Benjamin$^*$ et al. 2007 & WL & 0.78$\pm$0.05 & 4\\
Massey$^*$ et al. 2007 & WL & 0.91$^{+0.09}_{-0.07}$ & 5\\
Gladders et al. 2007 & CL & 0.67$^{+0.18}_{-0.13}$ & 6\\
Seljak et al. 2006 &  Ly$\alpha$+CMB & 0.85$\pm$0.02 & 7 \\
& +CL+SN & \\
\hline
\end{tabular}
\end{center}
$^*$ assuming $\Omega_m$=0.27\\
1, 2: \cite{wmap}
3: \cite{rozo}
4: \cite{ben}
5: \cite{massey}
6: \cite{gladders}
7: \cite{seljak}
\label{table:sig8}
\end{table}







The motivation for this paper is the upcoming release of the peculiar velocity data from the Six Degree Field Galaxy Survey \citep[6dFGS,][]{6df}. 6dFGS has measured the redshifts of around 150 000 galaxies over almost the entire southern sky, with a subsample of around 12 000 galaxies having peculiar velocity measurements, an order of magnitude larger than any peculiar velocity survey to date. All previous peculiar velocity surveys \citep[eg ENEAR, Spiral Field I-Band (SFI) \& Mark III:][]{enear,sfi:gio,sfi:dacosta,markiii} have traced the velocity field only out to distances of around 7000 kms$^{-1}$ and suffered because of uneven sky coverage and the small number of galaxies. In addition to the new 6dFGS data an extended SFI sample \cite[SFI++,][]{sfi++1,sfi++} has recently been released consisting of around 5000 spiral galaxies. Furthermore the ever growing Type 1a supernova samples \citep[eg][]{jrk} mean that there is a wealth of data becoming available for peculiar velocity analysis.

Results from previous surveys which apply likelihood analysis \cite[see][for Mark III, ENEAR and SFI respectively]{zab97,zab,freud}  seemed to overestimate the combination $\sigma_8\Omega_m^{0.6}$ significantly compared to other probes at the time and to the current concordance cosmology.  Values of $\sigma_8$ from those analyses were in the region of 1.7 to 2.4 after assuming $\Omega_m=0.27$. 
Studies by \cite{hoffzar} and \cite{silb} show that this over-estimation may be due to inaccurate modelling of the nonlinear part of the power spectrum, the small scales.  This paper focuses on testing our ability to remove the bias from the nonlinear signal.

In this paper we develop a practical approach in which we bin the velocities on a grid and thus erase small scale information. Since this reduces the number of data points then it also makes a covariance matrix approach computationally feasible using the large amount of data from upcoming surveys.

The paper is organised as follows. In Section~\ref{section:pre} we introduce the radial peculiar velocity correlation function and likelihood analysis which the two methods used in this paper utilise in different ways. 
In Section~\ref{section:grid} we describe the binning method where the galaxies used in this analysis are binned on a grid. The effect of the size of the grid cells on the results is shown. 
In Section~\ref{section:pca} we overview the method of  \cite{om} and \cite{feldmanom} by using a more sophisticated principal components analysis (PCA) to remove the moments which are most sensitive to the nonlinear scales. 
In Section~\ref{section:mocks} we describe the mock catalogues used in testing the methods.  
The results are presented in Section~\ref{section:res}

\section{Preliminaries}
\label{section:pre}
The analysis used in this Section follows \cite{freud}, \cite{zab97} \& \cite{zab} and \cite{kaiser}. We use linear theory to predict the velocity correlation function and use a multivariate Gaussian to calculate the likelihood.

\subsection{Peculiar velocity correlation function}
\label{subsection:vcf}
To estimate cosmological parameters we compute the radial peculiar velocity correlation function (hereafter VCF) from linear theory for each parameter set.  This forms the basis for the prediction which will be compared to the peculiar velocity data using the likelihood function, see Section~\ref{section:like} below. The peculiar velocity with observational error can be represented as $\textbf{v}_i \cdot \hat{\textbf{r}}_i\equiv v_i =s_i+\epsilon_i$, therefore the observed VCF is defined by

\begin{eqnarray}
\label{eq:rij}
R_{ij}& =&\left< v_i v_j \right>=\left< s_i s_j \right>+\left<\epsilon_i \epsilon_j\right> \\
                           &=&\xi_{ij}+\epsilon_i^2\delta_{ij}
\end{eqnarray}and the average is over realisations of the universe. The first term is the signal VCF and the second term is the contribution from the errors in the velocity measurements. The errors are assumed to be uncorrelated so they only affect the diagonal terms in $R_{ij}$.  In linear theory, the signal part $\xi_{ij}$ can be split up into perpendicular and parallel components \citep{gorski,gjo}, which are scalar functions of r=$|\textbf{r}|$ 

\begin{equation}
\label{cf}
\xi_{ij}=\cos \theta_i\cos \theta_i\Psi_{||}(r)+\sin \theta_i\sin \theta_j\Psi_{\perp}(r)
\end{equation} where the angles are defined by $\cos\theta_{X}=\hat{\textbf{r}}_{X} \cdot \hat{\textbf{r}}$ and the diagonal elements $\xi_{ii}$ are given by Eq.~\ref{eq:diag} below. The $\Psi_{||}(r)$ and $\Psi_{\perp}(r)$ can be calculated from the matter power spectrum using linear theory and assuming all galaxies are approximately at redshift zero

\begin{equation}
\label{eq:psi}
\Psi_{||,\perp}(r)=\frac{H_0^2 f^2(\Omega_m)}{2 \pi^2} \int P(k) B_{||,\perp}(kr) dk
\end{equation}
where $B_{\perp}=j_0^{'}(x)/x$ and $B_{||}=j_0^{''}(x)$ and $ j_0^{'}, j_0^{''}$ are the first and second derivative of the zeroth order spherical Bessel functions respectively, $H_0$ is the Hubble constant and $\Omega_m$ is the density of matter in the universe normalised by the critical density, $f(\Omega_m)$ is  the derivative of the growth function.  The auto correlation is given by

\begin{equation}
\label{eq:diag}
\xi_{ii}=\frac{1}{3}\frac{H_0^2 f^2(\Omega_m)}{2 \pi^2} \int P(k)dk.
\end{equation}
The dependence on $\sigma_8$ enters through the normalisation of the power spectrum $P(k)$. The dependence on $\Omega_m$ enters through f$(\Omega_m)\cong\Omega_m^{\gamma}$, where $\gamma\simeq0.557$ (found from a fit to the growth function at redshift zero, \cite{wangstein}) and through its effect on the shape of the matter power spectrum. The power spectrum $P(k)$ is generated using CAMB \citep{camb}. The above equation for $f$ assumes that the galaxies are at low redshift. The full equation contains the growth rate at the redshift of each galaxy.  Peculiar velocity surveys using distance indicator relations (as 6dFGS does) such as the fundamental plane \citep{fp} or D$_n$-$\sigma$ \citep{dnsig} are unlikely to have data beyond a redshift of 0.05. The growth rate increases by less than 1 per cent between redshift zero and redshift 0.05 for a flat $\Lambda$CDM model with $\Omega_m=0.3$, so this is a good approximation for this paper. We also use the approximation that the Hubble expansion is constant, described simply by a constant expansion rate for all galaxies in the survey, when converting between distance and redshift in both the simulations and the likelihood analysis.

\begin{figure*}
\center
\epsfig{file=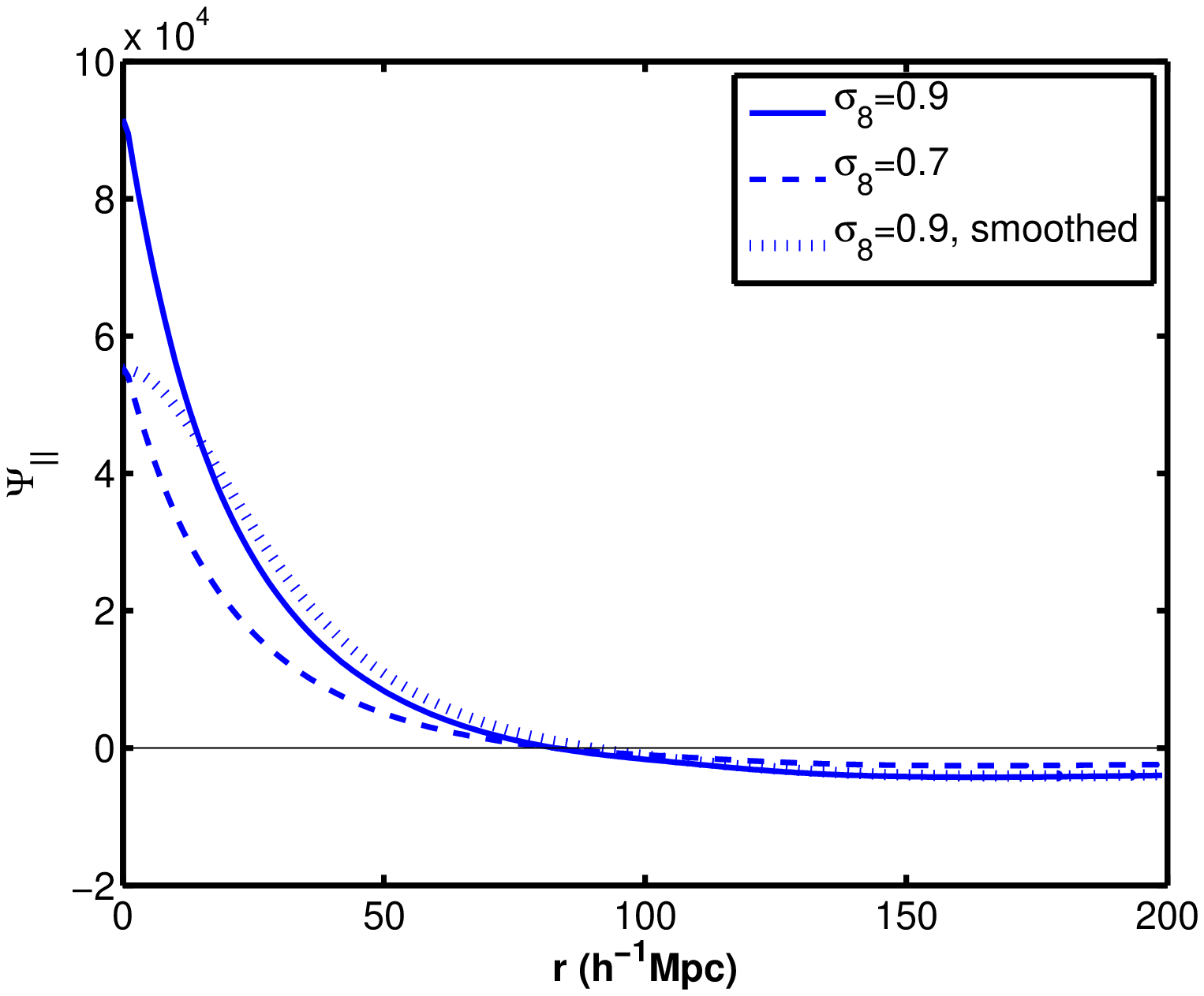,width=8cm,angle=0}
\epsfig{file=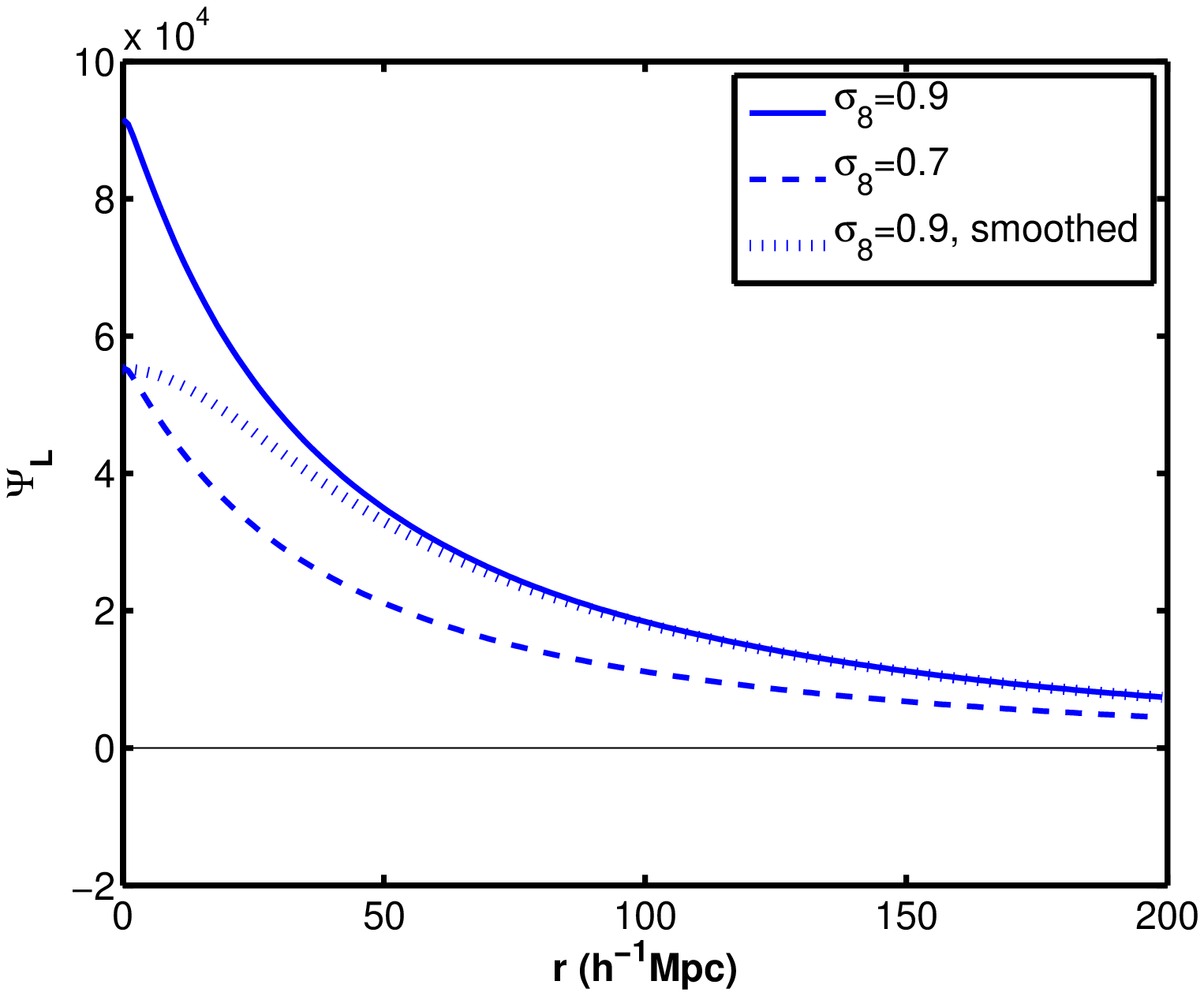,width=8cm,angle=0}
\caption{Left panel $\Psi_{||}$: the correlation between two galaxies which have their separation axis aligned exactly parallel to the line of sight. Right panel $\Psi_{\perp}$:  the correlation between two galaxies which have their separation axis aligned exactly perpendicular to the line of sight.  The solid line represents $\Lambda$CDM with $\sigma_8$=0.9, the dashed line represents  $\Lambda$CDM with $\sigma_8$=0.7, the dotted line represents $\Lambda$CDM with $\sigma_8$=0.9 and where the functions smoothed on a scale of 25$h^{-1}$Mpc. Note the closer similarity between the functions with the lower value of $\sigma_8$ (dashed) and the higher $\sigma_8$ + smoothing (dotted), than when the values of $\sigma_8$ are the same but one function is smoothed (dotted and solid). }
\label{fig:psi}
\end{figure*}

The two component correlation functions $\Psi_{||}$ and $\Psi_{\perp}$ are illustrated by the solid lines in Figure \ref{fig:psi}, respectively.
They represent the correlations between two galaxies which have their separation axis aligned exactly parallel ($\Psi_{||}$) or perpendicular ($\Psi_{\perp}$) to the line of sight.  As expected the correlation between the velocities of two galaxies decreases with increasing separation: the further apart they are the more likely different potentials are the main source influencing their motion.  It is interesting to note the negative correlation in $\Psi_{||}$ after scales of about 75$h^{-1}$Mpc.   Two galaxies separated along the line of sight are likely to be composed of one in-falling and the other back-falling into the same over-density region, therefore having radial velocities of the opposite sign creating the negative correlation.   In general pairs of galaxies have contributions from each of these functions, the amount from each dependent on the angle between the separation axis and the line of sight, represented in Eq.~\ref{cf} by the sine and cosine functions.

\subsection{Likelihood analysis}
\label{section:like}

If $\textbf{m}$ is the vector of model parameters, e.g. [$\sigma_8$ $\Omega_m$ $H_0$], and \textbf{d} is the vector of N data points, then Bayes' theorem states that the posterior probability density of a model given the data is 

\begin{equation}
\textit{P}(\textbf{m} \mid \textbf{d}) = \frac{\textit{P}(\textbf{m}) \textit{P}(\textbf{d} \mid \textbf{m})}{\textit{P}(\textbf{d} )}.
\end{equation}  The denominator is a normalisation constant which we divide out in our analysis.  The probability of the data given the model, $\textit{P}(\textbf{d} \mid \textbf{m})$, is the likelihood function which we discuss below.

We may want to find the set of parameters $\bf\Theta_{max}=\left[\theta_1 ... \theta_N \right]$ that maximizes the posterior, and this is equivalent to finding the set that maximise the likelihood function, for the case of a uniform prior probability $\textit{P}(\textbf{m})$.  To calculate the covariance matrix $\bf{R}$ it is necessary to calculate the power spectrum for a given set of cosmological parameters $\bf\Theta_m$ (Section~\ref{subsection:vcf}). Assuming that the peculiar velocities and the observational errors are Gaussian random fields the likelihood function can be written as

\begin{equation}
\label{eq:like}
\textit{L}=\frac{1}{\sqrt{(2\pi)^N |\bf{R}|}} \exp \left( -\frac{1}{2} \sum_{i, j}^{N} v_i\,\, ({\bf R}^{-1})_{ij}\,\, v_j\right).
\end{equation} where $v_i$ are the observed radial velocities.  

\section{Methodology}
\label{section:method}
The methods described in this paper both use the peculiar velocity correlation function and likelihood analysis described above. This Section describes how we apply these techniques to the peculiar velocity data
in practice.

\subsection{Gridding Method}
\label{section:grid}
The gridding method is a way of averaging together the peculiar velocities of spatially close galaxies by laying a grid across the survey.  Averaging over a number of galaxies should allow the linear signal to dominate.  It is similar to a counts-in-cells approach used in galaxy surveys. We discuss a simple way to bin the velocity field and detail a practical approach to take account of the binning accurately in the VCF. The technique we implement here is designed to be simple and fast.

The likelihood analysis outlined in Section~\ref{section:like} above uses the individual galaxy peculiar velocities, $v_i$, as the data.  Determining cosmological parameters in this way does not take account of the nonlinear part of the peculiar velocity signal because, as stated above, we make our prediction for the VCF based only on linear theory.  The density field becomes nonlinear only on small scales, above a wave number ($k$) of about 0.2$h$ Mpc$^{-1}$.  

\label{section:gridmethod}
We lay down a grid across the survey and average together all the peculiar velocities within each grid cell so that
\begin{eqnarray}
\label{eq:cell}
v_m^{\prime} &=&\left<v_i \right>_{i\in m} \\
\epsilon^{\prime}_{m} &=& \frac{\left< \epsilon_i \right>_{i \in m}}{\sqrt{n_{m}}}
\label{eq:errorcell}
\end{eqnarray}
where $v_m^{\prime}$ is the radial peculiar velocity of the cell and $\epsilon_{m}^{\prime}$ is the error on the velocity of the cell; the angle brackets denote an average over all galaxies $i$ within the cell $m$. Here there is the approximation that the grid cells are small enough so the radial components to each galaxy within each grid cell are parallel. This is tested when we apply the method to both linear and nonlinear simulations and we do not see a significant problem from this approximation. Note that $\epsilon_i$ is the contribution to the correlation function from the random velocity errors of each galaxy and therefore the remaining contribution $\epsilon^{\prime}_m$ to the binned correlation function is reduced by the square root of the number of galaxies.

Averaging the data over a volume of space will essentially smooth the velocity field which is equivalent to damping the small scale contributions. This reduces the observed correlations because we average away some of the signal.  If the data is averaged on a grid and inserted directly into the equations in Section~\ref{section:pre} without accounting for the averaging in the correlation function then the cosmological parameters will be biased. This is illustrated in Figure~\ref{fig:psi}. The smoothed parallel and perpendicular correlation functions (dotted lines) are more similar to the parallel and perpendicular correlation functions with a lower $\sigma_8$ (dashed lines) than the unsmoothed functions with the same $\sigma_8$ (solid lines).

This type of binning is then taken account of by multiplying the power spectrum in Eq.~\ref{eq:psi} and~\ref{eq:diag} with a window function corresponding to the size and shape of the grid cell.  This is because the binning in real space can be written as a convolution with a kernel followed by a sampling at the bin centres. The convolution kernel $W({\bf x})$ is uniform within a bin centered on the origin, is zero outside and normalised to have unit integral. In Fourier space this convolution is simply a multiplication, and the 
Fourier space window function is the Fourier transform of the real space window function 

\begin{equation}
W(k)=\left< \frac{8}{L^3}\frac{\sin \left(k_x \frac{L}{2}\right)}{k_x}\frac{\sin \left(k_y \frac{L}{2}\right)}{k_y}\frac{\sin \left(k_z \frac{L}{2}\right)}{k_z}
\right>_{{\bf k} \in k}
\end{equation}
where L is the length of a side of a cell in the grid, and the angle brackets denote an average over Fourier space directions.  This means equations that Eq.~\ref{eq:psi} and~\ref{eq:diag} become
\begin{eqnarray}
\label{eqn:sismooth}
\Psi^{\prime}_{||,\perp}(r)&=&\frac{H_0^2 f^2(\Omega_m)}{2 \pi^2} \int W^2(k)P(k) B_{||,\perp}(kr) dk \\
\label{eqn:sidiagsmooth}
\xi^{\prime}_{mm}&=&\frac{1}{3}\frac{H_0^2 f^2(\Omega_m)}{2 \pi^2} \int W^2(k)P(k)dk
\,
\end{eqnarray} where we use the position of the cell centre to calculate all the required distances. The corresponding VCF and likelihood are then formed using the smoothed quantities.

Unfortunately the method for accounting for the velocities described above assumes the data inside each grid cell is a continuous field, whereas it is in fact discrete values at the locations of the galaxies; we shall refer to this as the \textit{sampling problem}.  Galaxies trace discrete points of the peculiar velocity field, but if enough discrete points are averaged over then they will closely approximate averaging over a continuous distribution. However, some grid cells will not contain enough galaxies to provide a reasonable measure of the average of the velocity field within that cell, perhaps due to masked out areas in the survey or poor sampling in some areas.  The average radial velocity in a large volume of space tends towards zero; conversely when averaging over just a few galaxies in one cell the standard deviation of the velocities in that cell will be much higher than in well sampled cells.   If this effect is unaccounted for, the theory prediction will have to be larger to match the observed large velocity, and thus will bias $\sigma_8$ high.
Since the errors on the observed radial peculiar velocities are so large we find that in practice this bias is largely concealed by the noise, particularly as the error on each cell velocity is weighted by the number of galaxies in each cell, see Eq.~\ref{eq:errorcell}, so under-sampled cells are already weighted down by have larger noise than well sampled cells.

If we were to apply a truly accurate correction a separate window function would have to be calculated for every grid cell in the survey, and it would depend on the positions of the galaxies on the cell. This kind of calculation would be difficult and time-consuming. Instead we propse an approximate approach which we find to be sufficiently accuracte, as shown in Section~\ref{section:gridresults}.

This problem can be effectively corrected by reducing the size of the modification to the diagonal elements of the VCF, $\xi_{ii}$, according to the number of galaxies in each cell
\begin{equation}
\label{eq:corr}
\xi_{mm}^{{\rm corr}}=\xi_{mm}^{\prime}+\frac{\left(\xi_{mm}-\xi_{mm}^{\prime}\right)}{n_{m}}
\end{equation}
where $\xi_{mm}^{\rm corr}$ is the corrected value used to calculate the diagonal elements of the VCF. This correction uses the correct value for the diagonal elements of the correlation function in the limit that there is just one galaxy in the cell and also in the limit that there are infinite galaxies in the cell (a continuous field).  For very small numbers of galaxies in a large cell $\xi^{\prime}_{mm}$ is small compared to $\xi_{mm}$ and the velocities maybe relatively independent. In this case the diagonal elements decrease by the number of galaxies in the cell, by the same logic used to write Eq.~\ref{eq:errorcell}. For intermediate numbers of galaxies the value of diagonal elements of the VCF should be an improvement on using $\xi_{ii}^{\prime}$.  Applying this correction to just the auto correlations is reasonable since we find that they have the most power in constraining $\sigma_8$.  

Another popular method of grouping galaxies is a Friends-of-Friends algorithm \citep{fof}. Averaging groups identified by the Friends-of-Friends algorithm would result in a large range of group volumes due to the decrease in galaxy sampling with distance, unobserved fields in the survey and masked out areas due to the galactic plane or stars.  It would then be inappropriate to account for this smoothing with a window function of only one distinct length scale or shape. Instead a much more complicated method would have to be found.  The grid method of binning allows the problem to be tackled in the opposite way: it imposes a simple and regular volume average onto the survey, and therefore it is simple to account for the smoothing within the theoretical predictions.

\subsection{PCA method}
\label{section:pca}

This method was developed by \cite{om} and \cite{feldmanom} in the context of peculiar velocity analysis and uses Karhunen-Lo\`{e}ve methods of data compression \citep{kl1,kl2}.  A set of moments is created from the radial velocities which are insensitive to small scales and therefore to the nonlinearities.  These moments are then used in the likelihood analysis, see Section~\ref{section:like}. The moments $u_i$ are linear combinations of the line of sight peculiar velocities $v_j$ 
\begin{equation}
u_i=\sum_{j=1}^NB_{ij}v_j
\end{equation}
where ${\bf B}$ s a constant $N' \times N$ matrix, $N'$ is the number of moments retained and $N$ is the original number of galaxy velocities in the survey, so $N'  \le N$.  If $N'$ is less than $N$ there will be a loss of information, but by choosing the matrix ${\bf B}$ correctly this loss of information will be mostly associated with the velocity signal at small scales and therefore will tend to remove the effect of the nonlinearities.  

A simple model for the power spectrum is considered in which the power on nonlinear scales is proportional to a single parameter $\theta_q$. To be specific, we write the power spectrum as
\begin{equation}
P(k)=P_{\rm l}(k)+\theta_qP_{\rm nl}(k)
\end{equation}
where $P_{\rm l}(k)=0$ for $k> k_{\rm nl}$ and $P_{\rm nl}(k)=0$ for $k< k_{\rm nl}$.
To make the linear theory power spectrum $P_{\rm l}(k)$ we run CAMB using the same cosmological parameters as used to simulate the observed velocities.  Here $k_{\rm nl}$ is the wavenumber of the largest nonlinear density perturbation and we take this to be $k_{\rm nl}=0.2h$Mpc$^{-1}$.  We take $P_{\rm nl}(k)$ to be constant, $P_{\rm nl}(k)=P_0$, over the range of nonlinear scales, $k_{\rm nl}<k<k_c$.  This is not an accurate approximation to the nonlinear power spectrum however it has the benefit of greatly simplifying the calculations and essentially weighting all nonlinear scales equally.

The rms velocity in the presence of nonlinear effects is larger than the diagonal elements of the velocity correlation function predicted by linear theory. We write this difference as $\sigma_*^2$, as advocated in \cite{om}, and calculate it from noise-free nonlinear simulations. We use this difference to find a sensible value for $P_0$, while setting the fiducial value of $\theta_q$ to $\theta_q=1$. The relation between velocity dispersion and the power spectrum is used to show that
\begin{equation}
P_0=\frac{2\pi^2\sigma_*^2}{H_0^2 \Omega_m^{1.2}(k_{c}-k_{nl})}.
\end{equation} The choice of a maximum wavenumber $k_c$ reflects the fact that perturbations smaller than those which would form a galaxy will not contribute to a galaxy's velocity.  It can be shown that the method is fairly insensitive to the exact choice of $k_c$ and $k_{\rm nl}$ 
(the effect is of the order of a percent on $\sigma_8$).

If only a single moment is used as a data point then we can estimate the uncertainty on $\theta_q$ using the Fisher matrix formalism. The smaller the error on $\theta_q$, the more information is retained about small scales.  By finding the weightings $B_{nj}$ of the velocities $v_j$ which minimises the error on $\theta_q$
we can find the single moment that carries the minimum information about small scales. This problem is solved by introducing a Lagrange multiplier, formulating an eigenvalue problem and solving it to find $N$ orthogonal eigenvectors with corresponding eigenvalues, of the matrix ${\bf M}$ where
\begin{equation}
\label{eq:eigen}
M_{ij}=\sum_{i,j,m}L_{ki}^{-1}\frac{\delta R_{ij}}{\delta\theta_q}L_{lj}^{-1}
\end{equation}
and ${\bf L}$ is the Cholesky decomposition of the matrix ${\bf R}$, ${\bf R} = {\bf L}^T {\bf L}$. The significant contribution of the uncertainties to the diagonal terms makes the matrix especially well suited for decomposition.
The eigenvalue matrix, after sorting, can be shown to be equal to ${\bf LB}$ and thus ${\bf B}$ can be obtained after inverting ${\bf L}$ and applying it to the sorted eigenvalue matrix.

Keeping moments up to some $N'$, throwing away the moments with the largest eigenvalues, will ensure the data most affected by the small scales is lost. We find the number of moments to use, $N'$ using the Fisher matrix result 
\begin{equation}
\label{eq:momchoice}
\Delta\theta_q=\left(\frac{1}{2}\sum_{n=1}^{N'}\lambda_n^2\right)^{-1/2}
\end{equation} 
where $\lambda_n$ are the eigenvalues. Since our true value of $\theta_q$ is $\theta_q=1$ we desire to have enough modes so we cannot distinguish $\theta_q$ from zero. We require this at 1$\sigma$ confidence following \cite{om}, thus $\Delta \theta_q=1$. 

The moments $u_i$ are statistically uncorrelated by design, if we can assume that the peculiar velocities $v_j$ are Gaussian random variables. This means information contained by the moments thrown away will be completely removed from the data. See \cite{om} for full details. The likelihood function then becomes
\begin{equation}
L=\frac{1}{\sqrt{2\pi^N |{\bf \tilde{R}}|)}}\exp \left( -\frac{1}{2} \sum_{i, j}^{N'} u_i  ({\bf \tilde{R}}^{-1})_{ij} u_j  \right)
\end{equation}
where ${\bf \tilde{R}}={\bf B R B^{-1}}$. 

\section{Mock Catalogues} 
\label{section:mocks}
We test the methods for peculiar velocity analysis described in this paper on simulations.  The cosmological parameters used in the simulations are as follows: $\sigma_8=0.9$, $\Omega_m=0.3$ where $\Omega_b=0.04$, $\Omega_\Lambda=0.7$, $H_0=70$kms$^{-1}$Mpc$^{-1}$ and $n_s=1$.  Two types of simulation were used, a simple linear theory simulation and a more realistic nonlinear N-body simulation.

\subsection{Linear theory simulations}
\label{section:linearsim}
Linear theory simulations were performed to check the accuracy of the approximate fix to the sampling problem. The galaxy positions were simulated to mimic the redshift distribution of the SFI survey, as described in Section~\ref{section:sfidist}. We simulate galaxy velocities with the correlation function ${\bf \xi}$ in Eq.~\ref{cf}. The number of galaxies in 6dFGS is so large that computation of the covariance matrix and eigenvalues was prohibitively slow. Therefore we verify the solution to the sampling problem on a smaller survey only.

\subsection{N-body simulations}
\label{section:nbody}
The ensemble of simulations used in the current work is a sub-set of the suite performed in \cite{warrenetal} using the Hashed-Oct-Tree code \cite[HOT,][]{salmonwarren,wands} a particle tree-code with periodic boundary conditions. More specifically, the self-gravity of the system of dark matter particles is evolved according to
\begin{equation}
\frac{d ^2\vec{r_i} }{ d t^2} = \sum_{j=1} ^{Npart.} \frac{ G m_j \left(  \vec{r}_j -   \vec{r}_i \right)}{ \left(   \left | \vec{r}_j   - \vec{r}_i \right |^2 + \epsilon ^2 _{\mbox{\scriptsize{grav.}}}  \right)^{3/2} } + H^2(z)\Omega_\Lambda \vec{r}_i,
\end{equation}
where $\vec{r}_i$ and $m_i$ are the position and the mass of the $i$-th particle respectively. The Hubble Constant for a flat $\Lambda$CDM model at some redshift $z$ is  $H^2(z)\equiv H_0 \left[  \Omega_m(1+z)^3+\Omega_\Lambda  \right]$, with $H_0= 100h$kms$^{-1}$Mpc$^{-1}$ and $\Omega_\Lambda$ 
is the dimensionless cosmological constant at the present epoch. The second term on the {\it r.h.s} expresses an acceleration with respect to position of the $i$-th particle in the presence of the nonvanishing  cosmological constant. 
The initial conditions were set up for a flat geometry universe with parameters as stated as at the start of this section.

\subsection{Making Mock Catalogues}
\label{subsection:makingmocks}

To extract a generic 6dFGS mock catalogues we used three $1012^3$ particle simulations in $L_{\mbox{box}} = 384 \mbox{Mpc}h^{-1}$ boxes. The mass of the particles is $4.39\times 10^{9} M_{\odot}$ and the softening length $12.3 \mbox{kpc} h^{-1}$.  The following prescription to draw the catalogues from the above simulation was implemented:
\begin{itemize} 
 \item{ }
Using a Friends-of-Friends (FOF) method \citep{frenk} we identified agglomerates of dark matter particles. To be a bona-fide halo each agglomerate should have a minimum of 400 particles \citep{warrenetal}  within an isodensity  surface generated by a linking parameter $b \sim 0.2$ times the average interparticle separation in the simulation.   The halo velocity is the average velocity of the dark matter particles within a halo;
 
\item We split each of the three simulation boxes into eight sub-volumes of roughly the same size. To mimic the 6dFGS geometry (the southern sky sampled within $\approx 170h^{-1}$Mpc )
in each sub-volume we implemented the following iterative sequence: 
 \begin{itemize}
 \item{   } we chose three orthogonal directions and a halo randomly;
 \item{   } we carved a 6dFGS volume within the considered sub-volume having the chosen halo as an ``observer''  and the random orthogonal directions as the mock system of coordinates.  We returned to the previous step if this procedure was not possible;
\item{  } we computed the velocity field components  to take into account the rotation from the simulation's frame to the new frame defined by the three random orthogonal directions. We went  back and started the sequence on a new sub-volume. 
 \end{itemize}
 On average, the number density of halos in the simulations is $\sim 0.0043$~$h^{3}$Mpc$^{-3}$~while the 6dFGS number density  is  $10^{-2.126}$~$h^{3}$Mpc$^{-3}$ in the $K$-band \citep{6dflf}. However, the total number of 6dFGS galaxies with measured distance is roughly a tenth of the total; 
  
\item{ } A Monte Carlo rejection was used to choose halos according to the 6dFGS observed radial distribution (Campbell et al, in preparation). 
\end{itemize}

Each mock contains the real space positions and velocities. Henceforth, in our analysis we will use 21 mocks. For the considered box size, $L_{\mbox{box}} = 384 \mbox{Mpc}h^{-1}$, one can draw 19.4 independent mocks:  6dFRS covers the Southern sky for $|b| > 10^{\circ}$ out to a distance of $170 \mbox{Mpc}h^{-1}$.  Thus the error bars calculated from the 21 mock catalogues are taken account of accordingly. For more details regarding the 6dFGS mock catalogues see Teodoro et al (in preparation).
\begin{figure}
\center
\epsfig{file=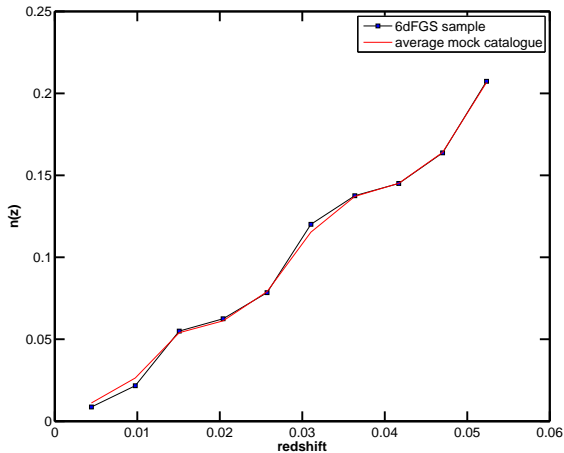,width=7.5cm,angle=0}
\caption{A comparison between the $n(z)$ of the mock catalogues and the $n(z)$ of the galaxies in the velocity sample of the 6dFGS survey. The black (dark) line with the square points is the $n(z)$ from a sample which the final 6dFGS sample will be drawn.  The red (light) line is the average $n(z)$ of the 21 mock catalogues.  Both have been normalised by the number of galaxies the samples contain.}
\label{fig:nofz}
\end{figure}
 Figure \ref{fig:nofz} shows a comparison between the $n(z)$ of the mock catalogues and the $n(z)$ of the galaxies in the velocity sample of the 6dFGS survey.

\subsection{SFI redshift distribution}
\label{section:sfidist}
Due to computing limitations performing the full likelihood analysis using all the individual galaxies can only be done for around one thousand galaxies.  To show the effect of binning compared to no binning we therefore mimic the galaxy distribution of an older survey which is smaller than the 6dFGS survey we are primarily interested in.  For this purpose we roughly recreate the $n(z)$ from the SFI survey, which contains 1289 galaxies.  The locations of the galaxies were simulated by picking 1289 galaxies to follow the SFI $n(z)$ from the mock catalogues described in Section~\ref{section:nbody} above.  The angular distribution of the galaxies was selected at random
from the full survey area of sky of the 6dFGS mock catalogues (roughly a hemisphere).
The same galaxy positions were used for both the linear and the nonlinear simulations. This was to ensure a similar sampling for the positions of the galaxies in the linear simulation to the nonlinear simulation. This is important because we are testing the accuracy of our fix to the sampling problem, which is related to the distribution of galaxies. 
 
\subsection{Velocity errors}
To measure the radial peculiar velocity of a galaxy, the redshift and distance to that galaxy first need to be measured. The velocity can then be found from $v_{pec}=cz-d$.  It is the measurement of the distance $d$ that is so difficult because the scaling relations between galaxy properties that are used to estimate the distance have a very large scatter.  They are also subject to a number of biases, of which Malmquist bias \cite[see][for a review]{S&W} is the most difficult to correct.  The effect of Malmquist bias is also to over-estimate $\sigma_8$ and $\Omega_m$ which is the same direction as the effect of nonlinearities. Since the aim of this paper is to focus on quantifying how well the analysis removes the effect of the nonlinear signal, we take the idealised situation that we have correct the Malmquist bias perfectly.  To replicate the effect of the scatter in the distance indicators we added a Gaussian random error of width 20\% of the galaxy distance to each peculiar velocity.  The same procedure was carried out in \cite{feldmanom} when testing their optimal moments method described above.  

This idealistic treatment of the velocity errors is not unreasonable because Malmquist bias can in principle be corrected if the line of sight density distribution is known.  To correct for Malmquist bias in the real world usually the galaxy density distribution, or $n(r)$, from a different galaxy survey is utilised.  The correction only works if the galaxy survey used to compute the Malmquist bias corrections has the same underlying $n(r)$ as the peculiar velocity survey. To mimic this situation for the analysis methods in this paper another galaxy distribution would have to be created from the mock catalogues to be used for the Malmquist bias correction and therefore the success of the correction would depend on the properties of this simulation, which is beyond the scope of this paper.

\section{Results}
\label{section:res}
Here we present our results from using the methods described in Section~\ref{section:method}.  The aim is to show the ability of the methods to constrain $\sigma_8$ so that the result is not biased by nonlinearities.

\subsection{Results from the Gridding Method}
\label{section:gridresults}

\begin{figure}
\center
\epsfig{file=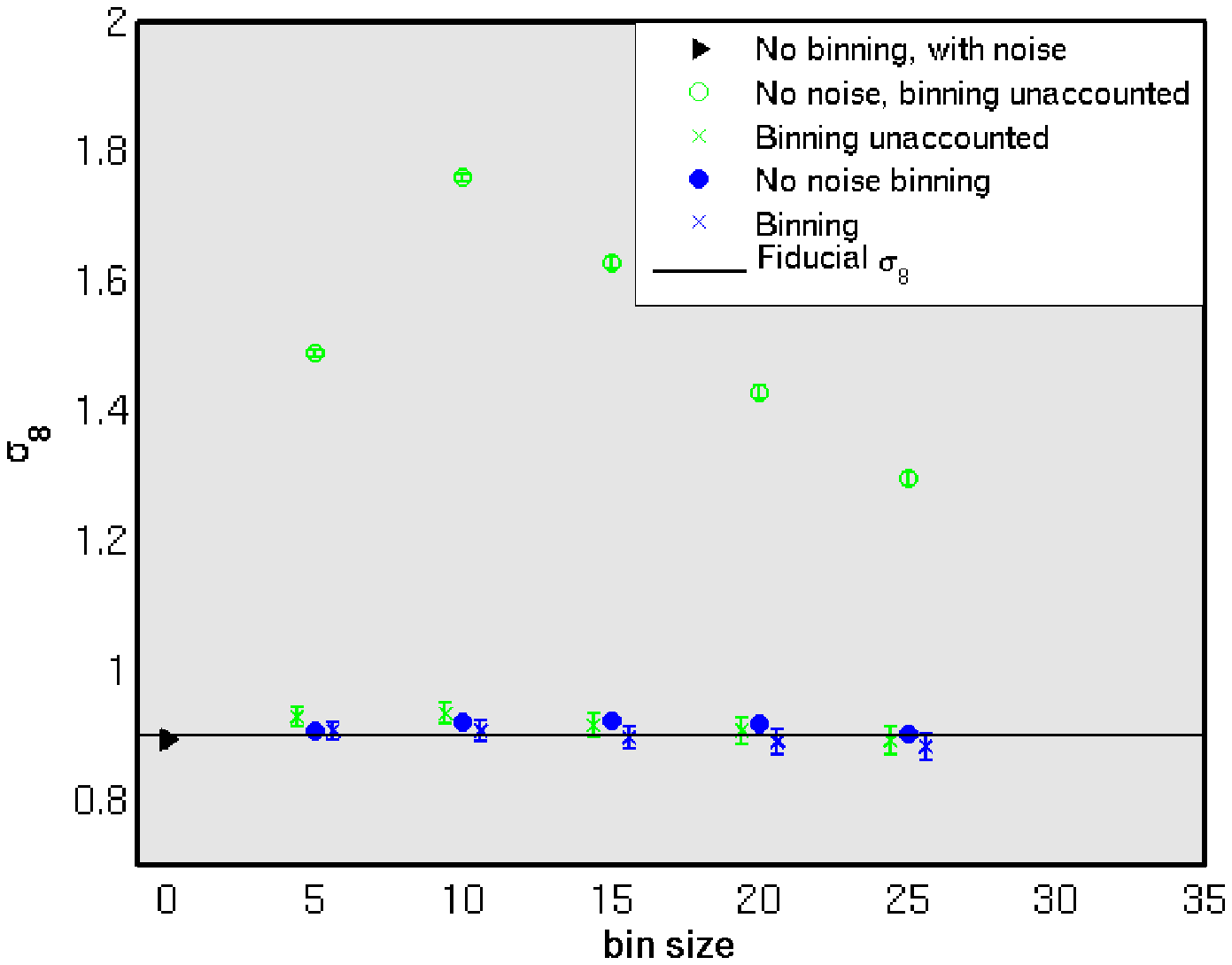,width=7.5cm,angle=0}
\epsfig{file=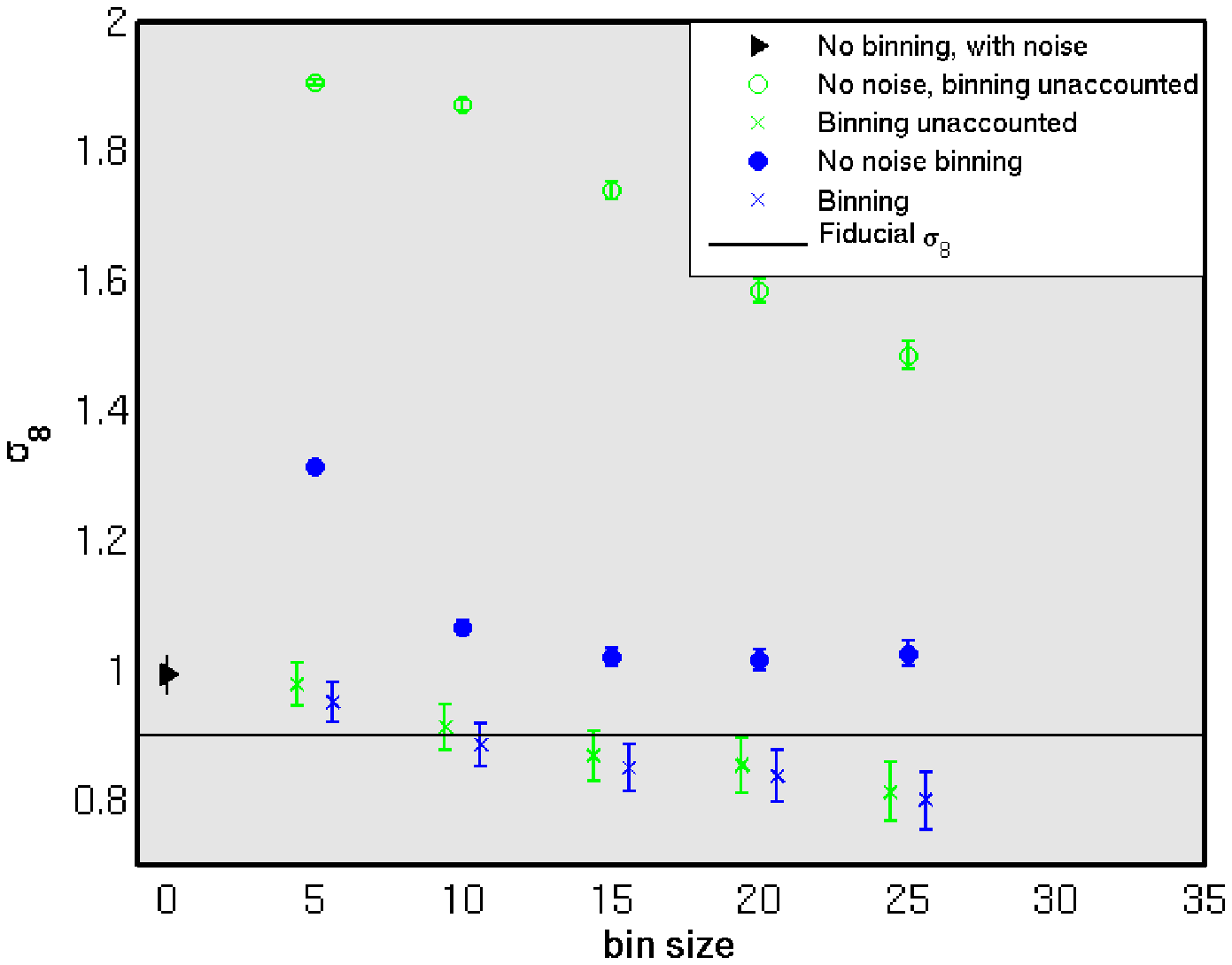,width=7.5cm,angle=0}
\epsfig{file=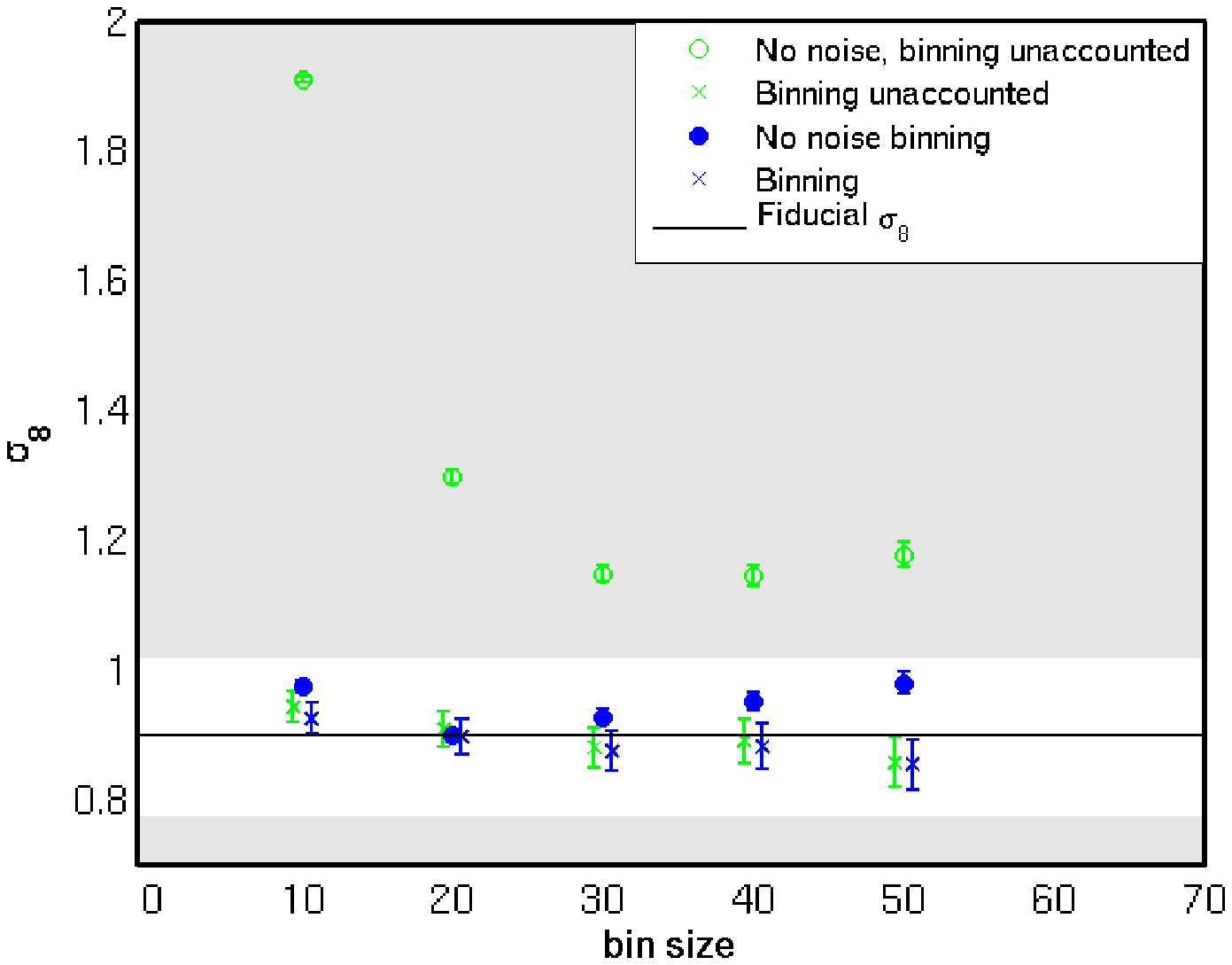,width=7.5cm,angle=0}
\caption{The green cross and green circle points are the results uncorrected for the sampling problem with and without uncertainties respectively. The blue cross and blue circle points are the results corrected for the sampling problem using equation \ref{eq:corr}, again with and without uncertainties respectively. The black triangle point in the top and middle panel is the result from doing the analysis with all the individual galaxies.  The error bars in the middle and bottom panels are the mean 1$\sigma$ error on $\sigma_8$ from each mock divided by the square root of the number of independent mocks (19). Top panel: Linear simulation with 1289 galaxies.  The galaxy positions were selected from the 6dF mock catalogues to closely resemble the $n(z)$ of the SFI survey.
Middle panel: Nonlinear simulation with 1289 galaxies. Galaxies were selected from the 6dF mock catalogues to closely resemble the $n(z)$ of the SFI survey. 
Bottom panel: Nonlinear simulation with 13000 galaxies selected from the 6dF mock catalogues. The white area depicts clearly the expected constraint on $\sigma_8$ from a bin size of 20Mpc/h, since the errorbars shown are divided by the square root of the number of independent mock catalogues}
\label{fig:gridding}
\end{figure}

We applied the method in three different situations (see Section~\ref{section:mocks}): to a linear simulation of a small survey, a nonlinear simulation of a small survey and a nonlinear simulation of the 6dFGS survey.  The results are shown in Figure~\ref{fig:gridding}.

The top panel of Figure~\ref{fig:gridding} shows the results from using a linear theory simulation of a 
survey with 1289 galaxies and a SFI-like $n(z)$.  All cosmological parameters are kept fixed except for $\sigma_8$, for which we calculate 68 per cent confidence limits. We take the galaxy positions from an N-body mock catalogue, to incorporate the galaxy clustering, and perform 100 velocity noise and galaxy position realisations. The points shown are the average values of $\sigma_8$ across these. The first point, at a bin size of zero, is the result of a standard likelihood analysis as described in Section~\ref{section:pre}, using all the galaxies separately. As expected, the point agrees well with the fiducial model.

The light (green) circles show results from noise-free simulations in which we have binned the velocity data but treated these binned velocities as separate independent tracers of the point velocity field in the theoretical predictions, i.e. we have used Eq.~\ref{eqn:sismooth} and \ref{eqn:sidiagsmooth} instead of the approximate correction for the sampling problem on the diagonal elements, given by Eq.~\ref{eq:corr}. We see that the results are strongly biased, since there are some cells containing only a very small number of galaxies. In these cells the velocity field has not been smoothed considerably, yet the theoretical prediction in Eq.~\ref{eqn:sidiagsmooth} has been smoothed, so $\sigma_8$ has to be large to fit the data. The results are shown for different bin sizes, and we see that as the bin size is increased the smoothed predictions (Eq.~\ref{eqn:sidiagsmooth}) work increasingly well, as expected.
At small bin sizes we also expect the results to be more accurate because the data and theory are both smoothed very little.

The dark (blue) circles show results from the same noise-free data, in which we now use the sampling problem correction given in Eq.~\ref{eq:corr}. The fact that these are so close to the fiducial model is not trivial, since the correction is an approximate but practical one. The error bars are the survey errors divided by the square root of the number of realisations. Presenting the errors in this way does not represent the expected error from 6dFGS, but it shows whether or not  $\sigma_8$ is biased significantly. The points agree with the fiducial within the error bars. It would be necessary to perform more simulations to test the accuracy of the sampling problem correction further, and ultimately we expect it to break down. These simulations tell us that the correction works to better than $2.5$ percent on $\sigma_8$, which is sufficient for the near future velocity data. 

The crosses show results from simulations in which noise is added to each velocity at the 20 per cent level. The light (green) crosses are the results when the sampling problem correction is not implemented. Whereas the lack of this correction caused a very large bias on $\sigma_8$ for the noiseless velocity simulation, we see that the bias is not significant when a realistic noise level is used.
The dark (blue) crosses correspond to realistic noise simulations with the sampling problem correction implemented. These points are very slightly lower than those without the correction included (note that the same noise realisations were used for each so this relative difference is significant despite the size of the error bars). But the difference is small. Since there is so little difference between the blue and green cross points we conclude that doing an even more accurate window function involving the exact positions of every galaxy in each grid cell would be unnecessary as it would offer negligible improvement.
Finally because the fiducial is recovered beautifully by all but the uncorrected no noise points we find that the approximation mentioned in Section~\ref{section:grid}, that the radial components of all galaxies in each grid cell are parallel, is good enough.

The middle panel of Fig.~\ref{fig:gridding} shows the effect of nonlinear theory, serving as a direct comparison with the top panel. The same survey parameters are used as for the top panel, and the only difference is the use of the nonlinear theory for the simulated noise-free velocities. The mock catalogues described in Section~\ref{section:mocks} are used, a total of 21 different catalogues with the same fiducial cosmology. Again the point at zero bin size corresponds to the traditional method in which all galaxies are treated independently. This is expected to be biased by nonlinear theory, and indeed we see that the difference from the fiducial model is significant and around 10 percent. This is not a large bias, but we have not matched the SFI sky coverage in our simulation, which would have been smaller and thus more susceptible to nonlinearities.

The remaining points correspond directly to their counterparts in the top panel. The biases on the noise-free points are larger than before, especially at small bin sizes. Even for the corrected noise-free points (dark/blue circles) there is a disagreement which persists even to large bin sizes.
The distribution of galaxies in the top panel of Figure~\ref{fig:gridding} 
exactly matches the distribution in the middle panel. The difference between the top and middle panels is just the peculiar velocity values themselves.  In the top panel the peculiar velocities are derived from a linear theory simulation and in the middle panel they are derived from a nonlinear N-body simulation.  As can be seen from the top panel the fiducial $\sigma_8$ is recovered precisely for all bin sizes and all points except the wholly unrealistic uncorrected no noise points.  It should be noted that the no noise points with the sampling correction do recover the fiducial.  This shows that the method by itself does not cause any systematic uncertainties on the $\sigma_8$ result, and shot noise and velocity errors are properly accounted for.

The differences between the top and middle panels of Figure~\ref{fig:gridding} are therefore entirely due to the realities of the N-body simulation including nonlinearities (expected on small scales) and cosmic variance (expected on large scales). Note that the cosmic variance effects will be larger for the SFI mock catalogues than for the 6dFGS catalogue since the number of selectable particles available in the finite number of simulations is smaller. This cosmic variance will be different for the noisy and noise free points because they sample different parts of the simulations (the noisy points effectively sample the very low redshift parts of the simulation only, whereas the no-noise points also sample higher redshift points). We do not address this further in this paper since our goal is to assess the contamination of  nonlinearities in 6dFGS-like data. 
To investigate these cosmic variance effects further is beyond the scope of this paper particularly since it would require a larger suite of simulations.
However this suggests that if a more accurate distance indicator, such as supernovae, were used in a survey with a similar geometry then this issue may need to be revisited.

The crosses in the middle panel of Fig.~\ref{fig:gridding} show the results of the small SFI-like simulation with a realistic noise level added to the N-body simulation velocities. The noise washes out the nonlinearities to a large extent, as seen by comparing points at a small bin size of 5 $h^{-1}$ Mpc, where the no-noise (corrected) dark (blue) circle is much higher than the points with a realistic noise level. The nonlinearities are expected to affect galaxies at small separations the most. The most extreme case of small separations is the velocity autocorrelation, i.e. the diagonal elements of the covariance matrix. The noise essentially blankets the predicted diagonals and therefore covers up biases due to nonlinear theory. Again there is little difference between the two sets of crosses, implying that the details of the sampling problem correction will not be extremely important on data with 20 per cent distance errors.
At large bin sizes the noisy points lie below the fiducial.  We attribute this to cosmic variance from the finite number of mock catalogues. Since it occurs when the bin size is 20$h^{-1}$Mpc or more, a volume which is at least a quarter of the size of the small survey, naturally the cosmic variance will be large.  We are unlikely to want to average over such large volumes in practice because we would loose too much information, therefore we do not obtain more mock catalogues to improve the result. The bias appears to have been removed for the point at 10 $h^{-1}$ Mpc, which is one-eighth of the survey size. 

The bottom panel of Fig.~\ref{fig:gridding} shows the results from using a nonlinear simulation of the 6dFGS survey with 13000 galaxies.  It is not possible computationally to do the analysis with a bin size less than 10$h^{-1}$Mpc, and at these scales the $\sigma_8$ result appears to be biased high at 68 per cent confidence. A bin size of 20$h^{-1}$Mpc appears optimum as there is a trade off between decreasing the constraint on $\sigma_8$ and overcoming the systematic bias due to the nonlinearities.
The resulting expected constraint on $\sigma_8$ from this bin size is depicted by the white area on the plot since the errorbars shown are divided by the square root of the number of mock catalogues.
At this bin size the bias on $\sigma_8$ from nonlinearities is less than 3 percent.  This is about one-fifth of the size of the error expected on $\sigma_8$ from 6dFGS at this bin size and therefore we conclude that we have devised a method suitable for removing the nonlinearities in 6dFGS.

\subsection{Results from the PCA method}

We applied the PCA method outlined in Section~\ref{section:pca} to two different survey types
(see Section~\ref{section:mocks}): to a nonlinear simulation of a small survey with a SFI-like galaxy distribution and a nonlinear simulation of the 6dFGS survey.  
We show results after averaging over all of the 21 mock catalogues.

\begin{figure}
\center
\epsfig{file=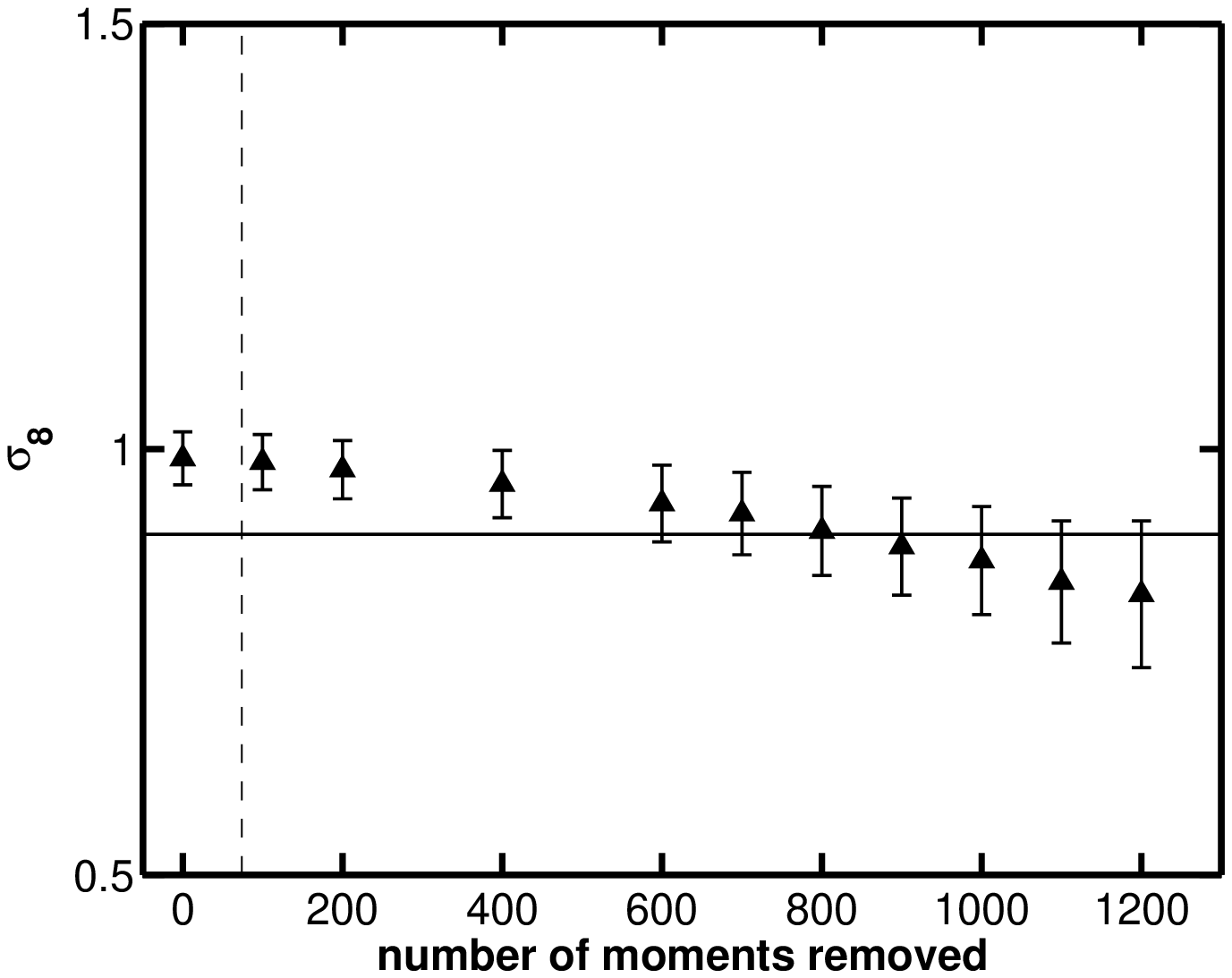,width=8cm,angle=0}
\epsfig{file=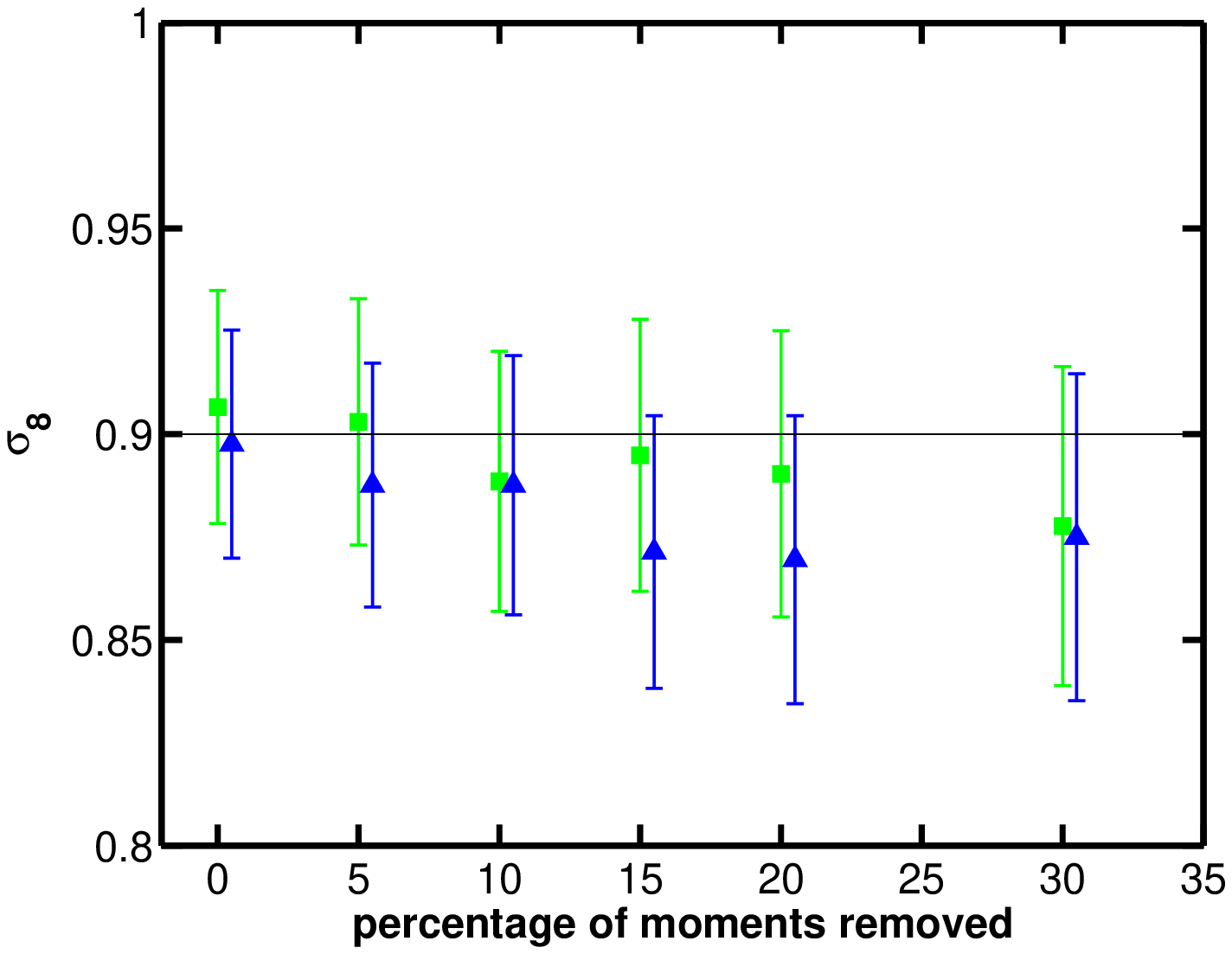,width=8.15cm,angle=0}
\caption{Top panel: The nonlinear simulation of 1289 galaxies with a SFI-like $n(z)$, just the points with measurement errors are shown. The vertical black dotted line shows the recommended number of moments to remove found from Eq.~\ref{eq:momchoice}.  Bottom panel:  The nonlinear simulation of the 6dFGS survey.  The points shown are with (dark/blue triangles) and without (light/green squares) the sampling problem correction applied. The horizontal solid line in both plots is the fiducial $\sigma_8$. The error bars in both panels are the mean 1$\sigma$ error on $\sigma_8$ from each mock divided by the square root of the number of independent mocks (19).}
\label{fig:pca}
\end{figure}

The top panel of Figure~\ref{fig:pca} shows the results from using a nonlinear simulation of 1289 galaxies with a SFI-like $n(z)$, with a realistic noise level. No binning is done to the data. 
The $x$-axis shows the number of number of PCA moments removed. Note that the point at zero moments removed is exactly equivalent to the point at bin size zero on the middle panel of Figure~\ref{fig:gridding}.
As expected, as more moments are removed the bias caused by nonlinear theory is reduced. The uncertainties also increase, as expected. The bias to low $\sigma_8$ values when a large amount of smaller scale information is removed is consistent with our cosmic variance explanation of the low points at large bin sizes for the middle panel of Fig.~\ref{fig:gridding}. 

The vertical black dotted line shows the recommended number of moments to remove found from Eq.~\ref{eq:momchoice}, and this number is about 74.  However, the bias has not reduced significantly at this point. It seems that a greater number of moments should be removed than recommended by Eq.~\ref{eq:momchoice}. When the bias on $\sigma_8$ disappears (at about 800 moments removed)
the error bar on $\sigma_8$ divided by the square root of the number of mocks is about 0.05.  Comparing this to the result for the smaller survey from the gridding method at the optimum bin size of 10$h^{-1}$Mpc (see middle panel Figure~\ref{fig:gridding}), the equivalent error is 0.03.  

The bottom panel of Figure~\ref{fig:pca} shows the results from using a nonlinear simulation of the 6dFGS survey with 13000 galaxies. Since we expect around 10,000 to 13,000 peculiar velocities from 6dFGS we cannot do the PCA analysis described above on the individual galaxy velocities.  This would mean doing calculations with at least 10,000x10,000 matrices which is too computationally intensive.
Instead for the nonlinear simulation of the 6dFGS survey we apply the PCA analysis to the data after it is binned with a cell size of 20$h^{-1}$Mpc using the gridding method (see bottom panel of Figure~\ref{fig:gridding}).

Results are shown with (dark/blue triangle points) and without (light/green square points) the sampling problem correction applied. This time the x-axis is percentage instead of the number of moments removed.  This is because after binning the data the total number of cells containing data varies with the original galaxy distribution.  Since the total number of cells varies between mock catalogues and realisations of mock catalogues to be sure of being consistent between realisations we instead remove the same percentage of moments each time.  This time Eq.~\ref{eq:momchoice} recommends that no moments need to be removed and this is borne out by the result in the bottom panel of Figure~\ref{fig:pca}.  
The PCA analysis in this particular case shows no improvement over the gridding method.


\subsection{Forecast for the 6dFGS survey}

\begin{figure}
\center
 \includegraphics[width=8cm]{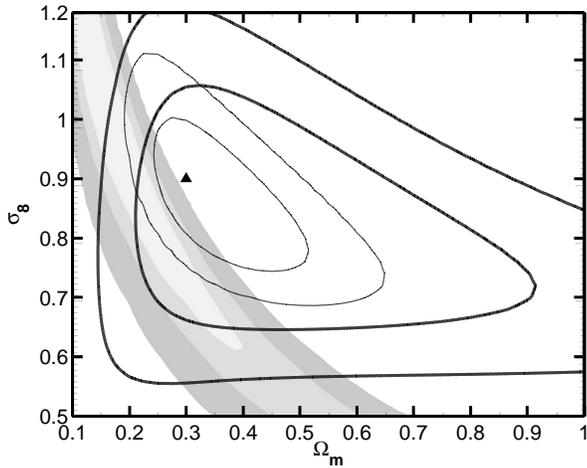}
\caption{The thick contours are the 1 and 2$\sigma$ contours for the gridding method with a bin size of 20$h^{-1}$Mpc when the sampling problem is corrected for with Eq.~\ref{eq:corr}. The thin contours are the same as the thick contours except repeated for a 5\% distance error, representing results from a supernova peculiar velocity survey. The fiducial $\Omega_m$ and $\sigma_8$ for the mock catalogues are indicated by the black triangle. The contours have been overlaid on the joint constraints on $\sigma_{8}$ and $\Omega_{m}$ from the 100 deg$^2$ weak lensing survey (grey scale contours) assuming a flat $\Lambda$CDM cosmology, depicting the 0.68, 0.95, and $0.99\%$ confidence levels.}
\label{fig:cont}
\end{figure}

From the nonlinear simulation of the 6dFGS survey we present 1 and 2$\sigma$ likelihood contours 
in the $\Omega_m$-$\sigma_8$ plane using the gridding method with a bin size of 20$h^{-1}$Mpc.  The resulting contours in Figure~\ref{fig:cont} are after averaging over all of the 21 mock catalogues.  
All cosmological parameters except for $\Omega_m$ and $\sigma_8$ were kept fixed.  The thick
contours are for the equivalent of the dark/blue cross point at 20$h^{-1}$Mpc in Figure~\ref{fig:gridding} (bottom panel), so the binning has been accounted for by using Eq.~\ref{eq:corr}. The fiducial $\Omega_m$ and $\sigma_8$ for the mock catalogues are indicated by the black triangle. After marginalising over $\Omega_m$ we find $\sigma_8=0.78\pm0.14$. 

The thin contours show the result for a peculiar velocity survey with an uncertainty that matches that expected from supernovae, of 5 percent error on the distance indicator, instead of the 20 percent assumed in the rest of this paper. 
The same 21 6dFGS mock catalogues are used except now the distance error has been reduced.  This is not a representation of a realistic future supernova survey since the sky and redshift coverage of such a survey would be much larger.  The resulting contours just show the increase in statistical power when using a distance indicator with the same observational error as supernovae.
The constraints on $\sigma_8$ at the fiducial $\Omega_m$ are a factor of two tighter from the supernovae than the galaxies. This is despite the factor of four decrease in velocity error, which implies that the cosmic variance limit is being reached. After marginalising over $\Omega_m$ we find $\sigma_8=0.84\pm0.08$.

The contours have been overlaid on the joint constraints on $\sigma_\mathrm{8}$ and $\Omega_\mathrm{m}$ from the 100 deg$^2$ weak lensing survey \citep{ben} assuming a flat $\Lambda$CDM cosmology and adopting the nonlinear matter power spectrum of \cite{smith}.  The weak lensing contours depict the 0.68, 0.95, and $0.99\%$ confidence levels. The models are marginalised, over $h=0.72\pm0.08$, shear calibration bias, and the uncertainty in the redshift distribution.

As can be seen from Figure \ref{fig:cont} the shape of the contours from the two different probes is different. This is because weak lensing and peculiar velocities probe the power spectrum differently.  The weak lensing contours follow lines of constant $\sigma_8\Omega_m^{0.68}$, the normalisation of the power spectrum, while the contours from peculiar velocities are more independent of the value of $\Omega_m$. $\Omega_m$ appears in the velocity analysis both in the normalisation of the velocity signal and in the shape of the matter power spectrum. The two effects partially cancel out to produce the flatter contours seen in the figure. The peculiar velocity contours close at low $\Omega_m$ but are more open at higher $\Omega_m$.

\section{Conclusions}
\label{section:conc}
In this paper we have demonstrated a new method, the gridding method, for the analysis of the 6dFGS survey that removes the contribution to the peculiar velocities from the nonlinear scales.  The method uses a grid to average over galaxies in the same volume of space, then performs the usual likelihood analysis with the averaging taken account for by multiplying the power spectrum in $\xi_{ij}$ with a window function describing the shape of the grid cells.   

We find a practical approach to taking into account the different effective amounts of smoothing in each grid cell, due to the different number of galaxies in each cell. We have shown that this works for linear theory simulations with and without noise. We find that it works well for nonlinear theory simulations for realistic amounts of noise. We find for a survey such as 6dFGS we can estimate $\sigma_8$ with less than 3$\%$ bias from nonlinearities, acceptable within the expected error bars.

We have compared our approach with the optimal moments method of \cite{om} and \cite{feldmanom} which uses a principal component analysis to select a set of optimal moments constructed from linear combinations of the peculiar velocities that are least sensitive to the nonlinear scales.  
We conclude that for our mock catalogues of the 6dFGS survey averaging in grid cells performs similarly well as compared with the PCA analysis. The error bar of the apparently unbiased point is a similar size for the PCA analysis as for the gridding method.  We have shown PCA is an excellent technique which advantageously uses data compression to successfully remove the nonlinear signal in the velocities.  However at some size of data set PCA cannot be applied to the whole survey, which could be overcome by dividing the survey up into smaller pieces because of the large matrices involved. This could be alleviated by performing the PCA analysis on each piece individually before combining the results although this looses some information on the largest scales in the survey, which are most accurately described by linear theory. Gridding is an alternative, though less elegant, data compression technique which retains the large scale information when dealing with large surveys.

In order to show the effect of the methods on removing biases from nonlinearities we did not consider other biases inherent in peculiar velocity analysis. In particular this paper did not attempt to deal with the minefield of Malmquist bias, arguably the next most difficult bias to correct; this issue will be
the subject of future work.

For the gridding method using a bin size of 20$h^{-1}$Mpc we presented the 1 and 2$\sigma$ likelihood contours for $\Omega_m$-$\sigma_8$ to show the constraint expected from 6dFGS.  The expected error on $\sigma_8$ after marginalising over $\Omega_m$ is approximately $16$ percent.  We also showed the constraints expected from a supernovae peculiar velocity survey the same size and sky coverage as the 6dFGS survey, this time the expected error on $\sigma_8$ is $0.08$ after marginalising over $\Omega_m$. 
A survey with this number of supernovae could potentially be performed using data from the GAIA
mission \citep{gaia}, which expects approximately 14 000 local Type Ia supernovae out to a redshift of $0.14$ over the whole sky. The different degeneracy direction in the $\sigma_8$-$\Omega_m$ plane will be useful in breaking degeneracies, see also \cite{GLS07}. 
Another survey commencing at the time of writing is the SkyMapper Telescope \citep{skymap} which aims to perform a multi-band survey of the southern sky and expects to discover 100 Type Ia supernovae per year out to a redshift of 0.085.
Furthermore, since velocities can probe $\sigma_8$ at very low redshift then they can in principle be combined with other probes to measure the growth of structure and thus properties of dark energy or modified gravity.

The constraints on cosmological parameters from 6dFGS will be a helpful addition to existing data, and we see that the accuracy is not greatly improved in using a more accurate distance indicator such as Type Ia supernovae. We conclude that the limitation is largely due to the small cosmic volume probed. Obtaining distance estimates to objects in a large volume is very difficult. The kinetic Sunyaev-Zel'dovich effect in clusters of galaxies is one possible method \citep{ksz} although this is a difficult challenge observationally \citep{khc,diaferio}. The current interest in large supernova surveys could potentially yield a useful data set, since the velocity error per unit area of sky is constant with distance.

The peculiar velocity data from the 6dFGS survey can be used to estimate not just $\sigma_8$ but also the shape of the matter power spectrum. This can be compared to results from the Two degree Field Galaxy Redshift Survey \cite[2dFGRS,][]{2df} and Sloan Digital Sky Survey \cite[SDSS,][]{sdss} which will help to constrain galaxy biasing.

\section{Acknowledgments}
John Lucey and Lachlan Campbell for providing some preliminary 6dFGS data.  Wolfram Freudling for providing the SFI redshifts.  Matthew Colless, George Efstathiou, John Huchra, Ofer Lahav, Alan Peel, Roman Scoccimarro, Andy Taylor, Saleem Zaroubi and Idit Zehavi for useful conversations.  AA acknowledges the receipt of a STFC studentship. SLB acknowledges support from the Royal Society in the form of a University Research Fellowship. LFAT acknowledges the financial support of the Leverhulme Trust.

\end{document}